%% file: itsc2018.tex
\documentclass[letterpaper, 10 pt, conference]{ieeeconf}
\pdfoutput=1

\usepackage{graphics}
\usepackage{epstopdf}
\usepackage{subfig}
\usepackage[export]{adjustbox}
\usepackage{multirow}

\usepackage{textcomp}
\usepackage{amsmath}
\usepackage{algorithm}
\usepackage{algorithmic}
\usepackage{mathtools}
\let\labelindent\relax
\usepackage{enumitem}
\usepackage{url}
\usepackage{color}

\title{\LARGE \bf
A Reinforcement Learning Approach to Jointly Adapt Vehicular Communications and Planning for Optimized Driving
}

\author{Mayank K. Pal, Rupali Bhati, Anil Sharma, Sanjit K. Kaul, Saket Anand and P. B. Sujit\\
{\tt\small{\{mayank15147, rupalib, anils, skkaul, anands, sujit\}@iiitd.ac.in}}\\
IIIT-Delhi, India
}

\newcommand{\state}[1]{x_{#1}}
\newcommand{\action}[1]{u_{#1}}
\newcommand{\actionlocal}[1]{u^{(l)}_{#1}}
\newcommand{\actionquery}[1]{u^{(q)}_{#1}}
\newcommand{\actionlocalset}{U^{(l)}}
\newcommand{\actionqueryset}{U^{(q)}}
\newcommand{\statelocal}[1]{x^{(l)}_{#1}}
\newcommand{\statequery}[1]{x^{(q)}_{#1}}

\begin{document}

\maketitle
\thispagestyle{empty}
\pagestyle{empty}

\begin{abstract}
Our premise is that autonomous vehicles must optimize communications and motion planning jointly. Specifically, a vehicle must adapt its motion plan staying cognizant of communications rate related constraints and adapt the use of communications while being cognizant of motion planning related restrictions that may be imposed by the on-road environment. To this end, we formulate a reinforcement learning problem wherein an autonomous vehicle jointly chooses (a) a motion planning action that executes on-road and (b) a communications action of querying sensed information from the infrastructure. The goal is to optimize the driving utility of the autonomous vehicle. We apply the Q-learning algorithm to make the vehicle learn the optimal policy, which makes the optimal choice of planning and communications actions at any given time. We demonstrate the ability of the optimal policy to smartly adapt communications and planning actions, while achieving large driving utilities, using simulations.

\end{abstract}
\input{introduction}
\input{related}

\input{problem}
\input{gridworld}
\input{qlearning}
\input{simulation}

\input{results}
\input{conclusions}

\bibliographystyle{IEEEtran}
\bibliography{itsc2018}

\end{document}

%% file: introduction.tex
\section{Introduction}
We consider an on-road environment that consists of autonomous vehicles equipped with sensors and other actors, for example, human driven vehicles, which may not have sensing ability. An autonomous vehicle would like to optimize its driving utility, for example, speed and/or smoothness of drive. To do so, it must choose a suitable motion plan while being cognizant of the other on-road actors. 

\begin{figure}
\begin{center}
\includegraphics[scale=0.4]{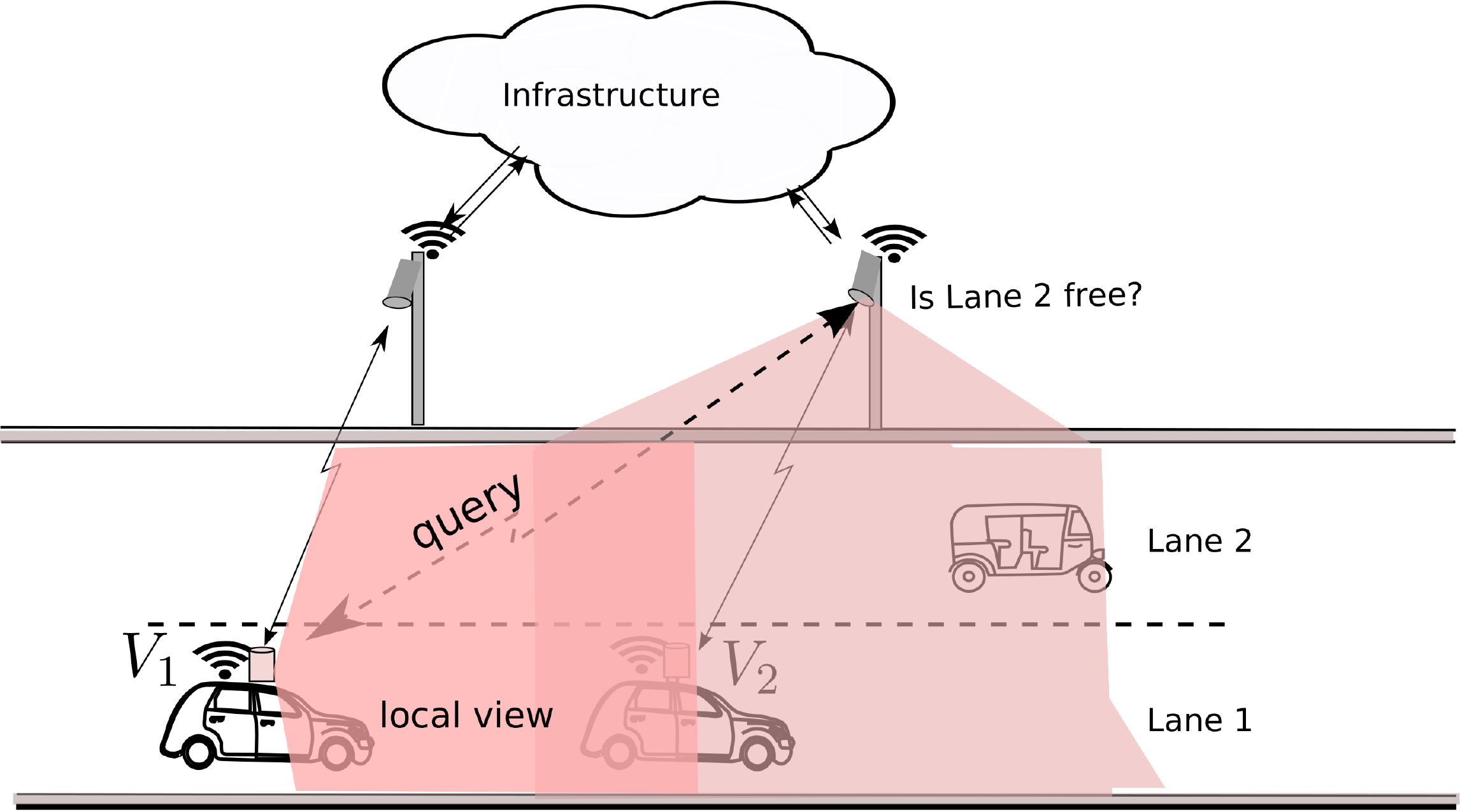}
\end{center}
\caption{Illustration of the on-road environment. A Vehicle may query the infrastructure for information that it uses to extend its local view. The infrastructure makes its own measurements and may receive measurements made by other vehicles too. The local view and the possible extended view is shown for the vehicle $V_1$.}
\label{fig:intro}
\end{figure}

Further, we envision the prevalence of roadside infrastructure sensors. Information sensed by vehicles or infrastructure may be communicated over a wireless network. In the absence of a network, an autonomous vehicle carries out motion planning given the road region it can perceive using its local sensors. Information from other sensors communicated over a wireless network allows the vehicle to perceive a larger region. Figure~\ref{fig:intro} provides an illustration. One expects this to allow for better motion planning and improved driving utilities.

However, \emph{communications} related \emph{constraints} impose limits on information that can be communicated at any given time. These limits impact the region that may be perceived by the autonomous vehicle and thus impact its motion plan. Conversely, an on-road environment, for example, a high vehicle density, may restrict the feasible motion plans and therefore diminish the requirement for communicating sensed information over the network.

In this work, we capture this interplay for a single autonomous vehicle (the ego vehicle) that uses the view perceived by its local sensors (\emph{local view}) and may choose to query information sensed by the infrastructure to append an \emph{extended view} to its \emph{local view}. The communications network is constrained and the ego vehicle may need multiple queries to populate its \emph{extended view}. The ego vehicle must adapt its \emph{motion plan} and its \emph{communications action} (specifically, the choice of sensed information queried from the infrastructure) to the on-road environment, such that its driving utility is optimized. We formulate this requirement as a reinforcement learning problem. 

Works that study the impact of communication constraints on driving utilities are limited. A recent work~\cite{commLimit} considers the limits on speed in autonomous vehicle networks that result from communications constraints. Works like~\cite{kimCP} and~\cite{Kim2013} consider techniques to merge locally sensed information with information obtained from other vehicles, to extend a vehicle's view. They use an $802.11$n/g wireless network as is for transferring sensed data amongst two to three vehicles. There are other works~\cite{delayMPatra}~\cite{ContentionFreeMAC_David_ICC2017} that don't dwell on the sensed information and motion planning. Instead, they consider problems related to vehicles sending information over a wireless network.

Our specific contributions are: 
\begin{enumerate}[leftmargin=*]
\item We formulate the problem of optimizing the driving utility of an autonomous vehicle using information obtained from its local sensors, and by querying the infrastructure over a network, as a reinforcement learning problem. Specifically, we formulate a discrete-time infinite horizon expected sum cost minimization. The goal is to find an optimal policy that the vehicle will use to choose its communications and motion planning actions at every time step.
\item We avail the occupancy grid representation of the on-road environment to capture both motion planning and communications related constraints. While simple, this helps us to clearly exhibit the interplay of planning and communications.
\item We use the Q-learning algorithm to enable the ego vehicle to learn the optimal policy, which solves the above problem, as it drives through a simulated on-road environment. Since Q-learning does not require the ego vehicle to know the model that guides evolution of the on-road environment, it can be used by an ego vehicle to learn optimal policies over time in real settings.
\item We compare the driving utility achieved by the ego vehicle, its choice of communications and motion planning actions when using the obtained policy, with alternate policies and under varied traffic densities and communications constraints. 
\end{enumerate}
The rest of the paper is organized as follows. In Section~\ref{sec:related} we summarize related works. In Section~\ref{sec:problem} we formulate the problem followed by details on the occupancy grid representation that we use in Section~\ref{sec:grid}. In Section~\ref{sec:qlearning} we explain how we use Q-learning to solve the problem. Section~\ref{sec:simulator} details the simulator. Section~\ref{sec:results} provides an evaluation of policies obtained using Q-learning for different scenarios. We conclude in Section~\ref{sec:conclusion}.

%% file: related.tex
\section{Related Works}
\label{sec:related}
There exists a large body of work that explores planning strategies for autonomous vehicles connected in a V2X framework. We focus on works relevant to our problem context, which involves cooperative perception and planning, communications in vehicular networks and applications of reinforcement learning to these problems. 

\emph{Cooperative Perception and Planning:} With increasing sensing and communication capabilities in cars, cooperative perception has become key to improving safety and traffic flow density. Several efforts \cite{kimCP, Kim2013, UncertainDist} have been made in extending the range of perception of the ego vehicle. Kim et al. \cite{kimCP, Kim2013} combine locally sensed information from various vehicles to create a merged occupancy grid map. The authors also study and characterize the effect of communication delay on map merging and conclude that safety critical tasks like collision avoidance should rely on local sensing information. They also suggest that longer-term decisions like early lane change or lane keeping could benefit from remote information. In \cite{UncertainDist}, the authors take a Bayesian approach to incorporate uncertainty in perception modules as well as communication delays in the merged occupancy grid, over which an appropriate algorithm like RRT* is applied for path planning. 

More recent works like \cite{Kamal_ITSC2017,Manzinger_ITSC2017} discuss cooperative perception and planning in mixed-traffic scenarios, where certain human driven vehicles may not have sensing and communication capabilities. 

Manzinger and Althof \cite{Manzinger_ITSC2017} develop an algorithm for cooperative collision avoidance by redistributing \emph{drivable regions} fairly among the cooperating vehicles. The human-driven vehicles in the occupancy maps are treated as any other obstacle and are assumed to be known at each time instance. Kamal et al. \cite{Kamal_ITSC2017} also work in a mixed-traffic, i.e., partially connected vehicle environment, and use the local and extended view to create a `road speed profile' of the upcoming road segment. The road speed profile is then used for anticipative planning and control to optimize a given utility function. 

These approaches develop the framework for merging the extended view with the local view and leverage the global view for better planning, however, all the schemes discuss broadcasting of information and do not consider the `usefulness' of the sensing information received by the ego vehicle. A na\"ive approach of collecting data may lead to unnecessary bandwidth consumption and processing delays.

\emph{Planning and Communications:} A crucial aspect of cooperative perception is the underlying protocol for communication and the constraints it imposes. In case of vehicular networks, often vehicles are equipped with multiple network interfaces like DSRC, Wi-Fi and Cellular. Higuchi and Altintas \cite{Higuchi_ITSC2017} explore hybrid schemes for vehicular communication and design a strategy for selecting one of the communication media (DSRC, Wi-Fi, Cellular, etc.) based on vehicular density on road regions. Talak et al. \cite{commLimit}, on the other hand, observe that any wireless communication scheme imposes constraints due to interference and delay and derive bounds for velocity as a function of traffic density. 

In \cite{Roth_AAMAS2005}, Roth et al. consider a multi-agent setting and learn a policy with each agent broadcasting its local view with no cost to communication. At execution time, however, a heuristic approach is taken, where an agent decides to communicate only when it benefits the team performance. More recently, Best et al. \cite{Best_ICRA2018} take a planning-aware communication approach for decentralized coordination of multiple agents. A particle filter framework is used where each agent tracks the action distribution of every other agent and decides to communicate with one only when its local utility can improve with new information.

%% file: problem.tex
\section{Model and Optimization Problem}
\label{sec:problem}
We model the interaction of the ego vehicle with its environment as a discrete-time dynamic system~\cite{bertsekas1995dynamic}
\begin{align}
\state{k+1} = f_k(\state{k},\action{k},w_k),\quad k=0,1,\ldots,
\label{eqn:stateEvolution}
\end{align}
where $k$ indexes discrete time, $\state{k}$ is the state observed by the ego vehicle at time $k$, $\action{k}$ is the action that the ego vehicle takes at time $k$, and $w_k$ is a random disturbance that captures the fact that the next state $\state{k+1}$, given the current state $\state{k}$ and action $\action{k}$, is governed by a probabilistic model. The function $f_k$ describes how the state is updated.

The state $\state{k} = [\statelocal{k}\ \statequery{k}]'$, for any $k$, is a $n\times 1$ vector. It consists of the two subvectors $\statelocal{k}$ and $\statequery{k}$. The former is a $n_l \times 1$ vector that is obtained from measurements made by sensors local to the ego vehicle, and the latter is a $n_q\times 1$ vector that the ego vehicle obtains by querying the infrastructure. The action $\action{k} = [\actionlocal{k}\ \actionquery{k}]'$ is a $2\times 1$ vector. Here $\actionlocal{k}$ is the \emph{motion planning} action that determines the vehicle's motion on the road (say, for example, its velocity or lane) for the following time step and $\actionquery{k}$ is the \emph{communications action} that determines the query the vehicle sends to the infrastructure. This query, at time $k$, populates elements in $\statequery{k+1}$.

Let $\actionlocalset(\state{k})$ and $\actionqueryset(\state{k})$, respectively, be the set of all motion related actions and communications related actions that are feasible in state $\state{k}$. While the set of feasible motion related actions restricts the on-road maneuvers the ego vehicle may make, the feasible set of communications actions captures the constraints on the ability of the ego vehicle to obtain information from the infrastructure using the communications network. An example of the former is a limit on the maximum acceleration or velocity and of the latter is a limited communications bandwidth/rate. For any $k$, $\action{k}$ is feasible only if $\actionlocal{k}\in \actionlocalset(\state{k})$ and $\actionquery{k}\in \actionqueryset(\state{k})$. 

At time $k$, given that the ego vehicle observes state $x_k$ and chooses action $u_k$, it incurs a bounded stage cost $g_k(\state{k},\action{k},\state{k+1})$. Let $X$ be the set of all states. Define a stationary (independent of time $k$) policy $\mu$, which is a function that maps every state $\state{k}\in X$ to a feasible action $\action{k}$. Let $J_\mu(\state{0})$ be the infinite horizon discounted expected sum cost when the ego vehicle starts in state $\state{0}$ and uses the policy $\mu$. In terms of the stage costs, we can write
\begin{align}
J_\mu(\state{0}) = E_{\state{1},\state{2},\ldots} \left[\sum_{k=0}^\infty \alpha^k g_k(\state{k},\mu(\state{k}), \state{k+1})\middle|\state{0}\right],
\label{eqn:sumCost}
\end{align}
where $0 < \alpha < 1$ is the factor that discounts future costs, and the expectation is over the sequence of states the ego vehicle visits, having started in $\state{0}$. In our work, we use $\alpha = 0.91$.

Our optimization problem is to find the optimal policy $\mu^*$ that minimizes the sum cost $J_\mu(\state{0})$, for all $\state{0}\in X$. Let $\Pi$ be the set of all feasible stationary policies. The optimal policy is 
\begin{align}
\mu^* &= \operatorname*{arg\,min}_{\mu\in \Pi}  J_\mu(\state{0}),\quad \forall \state{0} \in X.
\label{eqn:mustar}
\end{align}
The optimal policy $\mu^*$, at every time step, simultaneously chooses a motion planning and communications action, as a function of the state, such that the discounted sum cost in~(\ref{eqn:sumCost}) is minimized.

%% file: gridworld.tex
\section{Grid Representation}
\label{sec:grid}
\begin{figure}
\begin{center}
\includegraphics[scale=0.4]{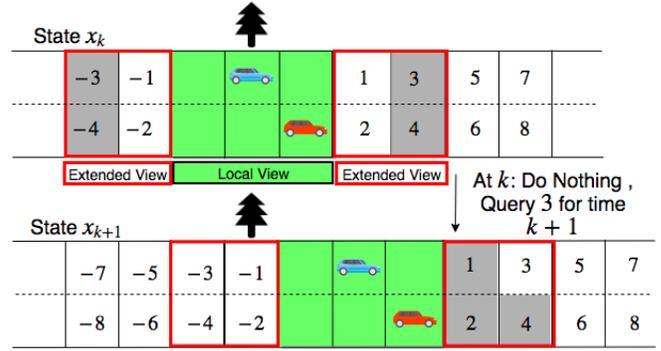}
\caption{The grid corresponding to the states $\state{k}$ and $\state{k+1}$ is shown. The ego vehicle (blue car) has a local view of six cells and an extended view of eight cells. Assume a velocity of $2$ cells/second. In $\state{k}$ one other car is in the local view. At any time $k$, the cells in the extended view are indexed relative to the local view of the ego vehicle. The occupancy of those with a grey background is unknown. The ego vehicle chooses to stay at a velocity of $2$ at $k$ and would like to query the cell that is indexed $3$ in its extended view at time $k+1$. This is cell $7$ in the grid at $k$. The control action $\actionlocal{k} = \text{Do Nothing}$ and the query is $\actionquery{k} = 3$.}
\label{fig:occupancyGrid}
\end{center}
\end{figure}
We use an occupancy grid to represent the on-road environment through which the ego vehicle is driving. Each cell in the grid is either empty or is occupied (has a stationary or a moving obstacle). The grid, relative to the position of the ego vehicle, is divided into the two regions of (a) \emph{local} view and (b) \emph{extended} view. The local region occupancy can be measured by the sensors local to the ego vehicle. On the other hand, the ego vehicle needs to query the infrastructure to obtain occupancy information in the extended view. Figure~\ref{fig:occupancyGrid} shows an illustration. In practice, cell occupancy information is obtained by using sensors like cameras and LIDARs that are observing the road region corresponding to the cell.

\emph{States in the grid representation:} The state $\state{k}$ at time $k$ is obtained from the occupancy grid as follows. The subvector $\statelocal{k}$ is constituted of the velocity of the ego vehicle, its lane, and the occupancy information of each cell that is in its \emph{local view} and can be measured\footnote{In this work we assume that the measurements are accurate. More generally, the state may also include measurement uncertainty.} by its local sensors. In Figure~\ref{fig:occupancyGrid}, assuming that the ego vehicle is moving at a speed of $2$ cells/second, and listing the cell occupancy in its local view (starting from the cell behind the ego vehicle and going anti-clockwise), we get $\statelocal{k} = [2\ \text{Left lane}\ F\ F\ F\ O\ F]'$, where $O$ and $F$ denote, respectively, an occupied and a free cell. 

The subvector $\statequery{k}$ consists of the occupancy of cells that are in the \emph{extended view} of the ego vehicle. Their occupancy is obtained by querying the infrastructure at $k-1$. The occupancy of a cell that has never been queried is set to $\text{Unknown}$, else it is set to the last queried value. In Figure~\ref{fig:occupancyGrid}, in state $x_k$, the occupancy of cells marked $-2,-1,1,2$ (relative to the ego vehicle (blue)) was queried earlier to be free and that of $-4,-3,3,4$ is unknown.

\textit{Actions in the grid representation:} In this work, we 
restrict the set $\actionlocalset(\state{k})$ of motion 
planning actions to $\{\text{Accelerate},\text{Decelerate},
\text{Do Nothing},\text{Change Lane}\}$. Accelerate and 
Decelerate correspond to the ego vehicle, respectively, 
increasing and decreasing velocity. Lane Change has the 
vehicle move to an adjacent lane. The action Do Nothing 
implies that the vehicle sticks to its current velocity and 
lane. Occupancy of one or more cells in the extended region at 
time $k+1$ may be queried together at time $k$. The set $
\actionqueryset(\state{k})$ includes actions, each of which 
corresponds to a group of cells, which will be in the extended 
view at $k+1$, being queried. The set also includes the action 
where in the ego vehicle chooses not to query. Larger numbers 
of cells in a group correspond to a larger available 
communications bandwidth (bits/sec). On the other hand, if 
groups contain fewer cells, in effect, the ego vehicle has 
access to a smaller communications bandwidth. In Figure~
\ref{fig:occupancyGrid}, the extended region has eight cells. 
Supposed we group them into the sets $R_1 = \{-1,-2,-3,-4\}$ 
and $R_2 = \{1,2,3,4\}$. To populate occupancy of either group 
at time $k+1$, the ego vehicle queries them at time $k$. We 
have $\actionqueryset(\state{k}) = \{\text{Query}\ R_1,\ 
\text{Query}\ R_2, \text{No Query}\}$. Implicitly, the 
communications bandwidth is $4$ cells/time step.

\emph{Uncertainty in the on-road environment:} We introduce uncertainty in the grid world by choosing a cell in the grid to be occupied with a certain probability $0 \le pOccupied < 1$, independently of the other cells in the grid, and unknown to the ego vehicle. As the ego vehicle moves, new cells are added and the occupancy of each cell is determined similarly. The ego vehicle gets to know the true occupancy of a cell once the cell is in its local view or if the cell is in the vehicle's extended view and it queries for its occupancy.

\emph{State evolution and costs:} The ego vehicle will learn the optimal policy $\mu^*$ using Q-learning, without an explicit knowledge of the state evolution model~(\ref{eqn:stateEvolution}), using state and action trajectories and stage costs obtained from a simulator. Next, we describe the Q-learning based method followed by a description of the simulator, where we describe the stage costs.

%% file: qlearning.tex
\section{Finding the Optimal Policy Using Q-Learning}
\label{sec:qlearning}
The Q-learning algorithm provides a simulation driven iterative method to find the optimal Q-factors $Q^*(\state{},\action{})$ for every state $\state{} \in X$ and action $\action{} \in \actionlocalset \times \actionqueryset$. To simplify notation, define set $U(\state{k}) = \actionlocalset(\state{k}) \times \actionqueryset(\state{k})$. The optimal Q-factor $Q^*(\state{},\action{})$ is the optimal infinite horizon expected sum cost when starting with the state-action pair of ($\state{}$,$\action{}$). It is related to the optimal cost $J_{\mu^*}(\state{})$ of starting in state $\state{}$ as per the following equation.
\begin{align}
J_{\mu^*}(\state{}) = \min_{u \in U(\state{})}Q^{*}(\state{},\action{}),\quad \forall \state{} \in X.
\end{align}
The optimal policy $\mu^*$ (Equation~(\ref{eqn:mustar})) is given by
\begin{align}
\mu^*(\state{}) = \operatorname*{arg\,min}_{u \in U(\state{})} Q^{*}(\state{},u),\quad \forall \state{} \in X.
\label{eqn:mustarQ}
\end{align}
Further define
\begin{align}
\widetilde{Q}(\state{k},\action{k},\state{k+1}) &= g(\state{k},\action{k},\state{k+1})\nonumber\\ &\quad+\alpha \min_{v \in U(\state{k+1})} Q(\state{k+1},v).
\end{align}

The Q-learning based approach is summarized in Algorithm~\ref{alg:qlearning}. It proceeds in an episodic manner. A given episode begins by randomly picking a certain initial state-action pair and running a long simulation trajectory ($stepsPerEpisode$ long) that starts at the state-action pair. For every episode, we provide the simulator a probability $pOccupied$ with which a cell in the grid must be occupied. For every state $\state{t}$ that the simulation trajectory visits, a motion planning and a communications action is picked uniformly and randomly, respectively, from the sets $\actionlocalset{(\state{t})}$ and $\actionqueryset{(\state{t})}$ of actions feasible in the state. Inputting these to the simulator gives the stage cost $g_t(\state{t},\action{t},\state{t+1})$ and the next state $\state{t+1}$. 

The Q-factor corresponding to the visited state and chosen action is then updated using the original Q-learning algorithm~\cite{bertsekas1995dynamic}. All other state-action pairs are left unchanged. In this update, $\gamma_t \in (0,1]$ is a step-size that when chosen appropriately guarantees the convergence of the Q-factors $Q(\state{},\action{})$ to the optimal Q-factors $Q^*(\state{},\action{})$. We set it to $0.01$.

For the simulation scenarios that we present later, we observed convergence for a maximum of $numEpisodes = 10^7$ and $stepsPerEpisode = 200$.

\begin{algorithm}[t]
\small
 \begin{algorithmic}
 	\STATE Inititalize: $Q(\state{},\action{})$, $\forall \state{},\action{}$, $numEpisodes$, $stepsPerEpisode$
 	\STATE Output: $Q^*(\state{},\action{})$, $\forall \state{},\action{}$.
	\WHILE{$episodeCount$ $<$ $numEpisodes$}
	\STATE $\state{0} \gets$ SIMULATOR(); \COMMENT{Get starting state from simulator}
	\STATE $pOccupied \gets$ Choose uniformly and randomly from $\{0,0.1,\ldots,0.8\}$;
	\STATE $t \gets 0$;
	\WHILE{$t$ $<$ $stepsPerEpisode$}
	\STATE $\actionlocal{t} \gets$ Choose uniformly and randomly from $\actionlocalset(\state{t})$;
	\STATE $\actionquery{t} \gets$ Choose uniformly and randomly from $\actionqueryset(\state{t})$;
	\STATE $u_t \gets [\actionlocal{t} \actionquery{t}]$;
	\STATE \parbox[t]{.5\textwidth}{$[\state{t+1}, g_{t}(\state{t},u_t,\state{t+1})]\gets$\\
	$\text{SIMULATOR}(\state{t},u_t, pOccupied)$;}
	\STATE $Q(\state{t},\action{t}) \gets (1 - \gamma_{t}) Q(\state{t},\action{t}) + \gamma_{t} \widetilde{Q}(\state{t},\action{t},\state{t+1})$\label{eqn:qUpdate};
	 \STATE $t \gets t+1$;
	\ENDWHILE		
	\STATE $episodeCount \gets episodeCount + 1$;
	\ENDWHILE
\end{algorithmic}
\caption{The Q-learning Based Algorithm}
\label{alg:qlearning}
\end{algorithm}

%% file: simulation.tex
\section{SIMULATOR}
\label{sec:simulator}
We found simulators like Flow~\cite{wu2017flow}, \emph{DeepTraffic}\footnote{\scriptsize{\url{https://selfdrivingcars.mit.edu/deeptraffic/}}} by MIT and the \emph{Auto Drive Simulator}\footnote{\scriptsize{\url{https://github.com/HugoTian/Auto_drive_simulation}}}. Flow integrates micro-traffic simulation and reinforcement learning to explore control strategies in various traffic scenarios. DeepTraffic and Auto Drive both use deep reinforcement learning techniques to optimize the motion plan of an ego vehicle. However, we were unable to find simulators that supported the functionality above together with the possibility of an extended view that can be exposed to the ego vehicle in response to its querying for information. Motivated by existing approaches we built a simple simulator using, amongst others, the Python libraries NumPy and Pygame. This enables us to grasp the effectiveness of policies that adapt motion and communications actions jointly. We will briefly describe how the different actions are simulated followed by a description of the stage costs.

The simulator keeps track of every cell in an occupancy grid that includes the local view and the extended view. The sizes of each view are fixed during a simulation. As the ego vehicle moves, at every time step, new cells are added and an equal number of cells leave the simulation. The simulator sets the occupancy of a new cell to \emph{Occupied} independently and randomly with the probability $pOccupied$ that is provided to it. The occupancy of the cells that leave are lost.

The ego vehicle has access to the occupancy of all cells in the local view. However, the occupancy of a cell in the extended view is revealed to it by the simulator only when the cell is queried.

The simulator maintains distance in the units of cells and, correspondingly, velocity and acceleration in the units of cells/time-step and cells/time-step$^2$.\footnote{For ease of presentation, we will skip the units.} The ego vehicle has a certain maximum velocity that can be set. The minimum velocity is $0$. For scenarios we evaluated, the cell size was set to that of the ego vehicle. Let $v_k$ be the velocity of the ego vehicle at the time step $k$ and let $a_k$ be the acceleration (a negative value implies deceleration) chosen by it. The number of cells $d_k$ covered by the ego vehicle and its velocity at time $k+1$ are calculated as
\begin{align}
d_k = v_k + \left \lfloor\frac{a_k^2}{2}\right \rfloor;\ 
v_{k+1} = v_{k} + a_{k}.\label{eqn:simDistVel}
\end{align}

To exemplify, if a vehicle moving at a velocity of $1$ at time $k$ decelerates at a rate of $1$, it will come to a halt at time $k+1$ in the cell it started in at time $k$. On the other hand, a vehicle that has a velocity of $0$ at time $k$, on choosing to accelerate at $1$, will have a velocity of $1$ at time $k+1$. However, it will be in the same cell as it was in time $k$.

Lastly, lane changes are executed by moving the ego vehicle by the number of cells given by~(\ref{eqn:simDistVel}) and then placing it in the corresponding cell in the chosen adjacent lane.

\subsection{Stage costs (negative rewards)}
To make the ego vehicle learn the optimal policy using Q-learning we must set the stage costs (see Equation~\ref{eqn:sumCost}) appropriately. We use a very simple reward structure. The ego vehicle accrues unit stage reward (stage cost of $-1$) for every cell it moves in its direction of travel. This is to ensure that the vehicle covers distance. To discourage unnecessary motion planning actions or communications actions, the ego vehicle gets a stage reward of $0.1$ when it chooses the motion planning action Do Nothing or a communications action of No Query. Note that a vehicle may choose both in a time step, in which case it will get a reward of $0.2$ added to the number of cells it covered. An example of an unnecessary motion planning action is that of changing lanes when all available lanes including the current are empty. Similarly querying the occupancy of a cell that is already known is an example of wasteful communications. When an action leads to an imminent collision, the vehicle receives a reward of $-1000$ (a very large stage cost). Its velocity is forced to $0$ and its position remains unchanged.

%% file: results.tex
\section{Evaluation and Results}
\label{sec:results}
\begin{figure}[t]
	\centering
		\includegraphics[scale=0.35]{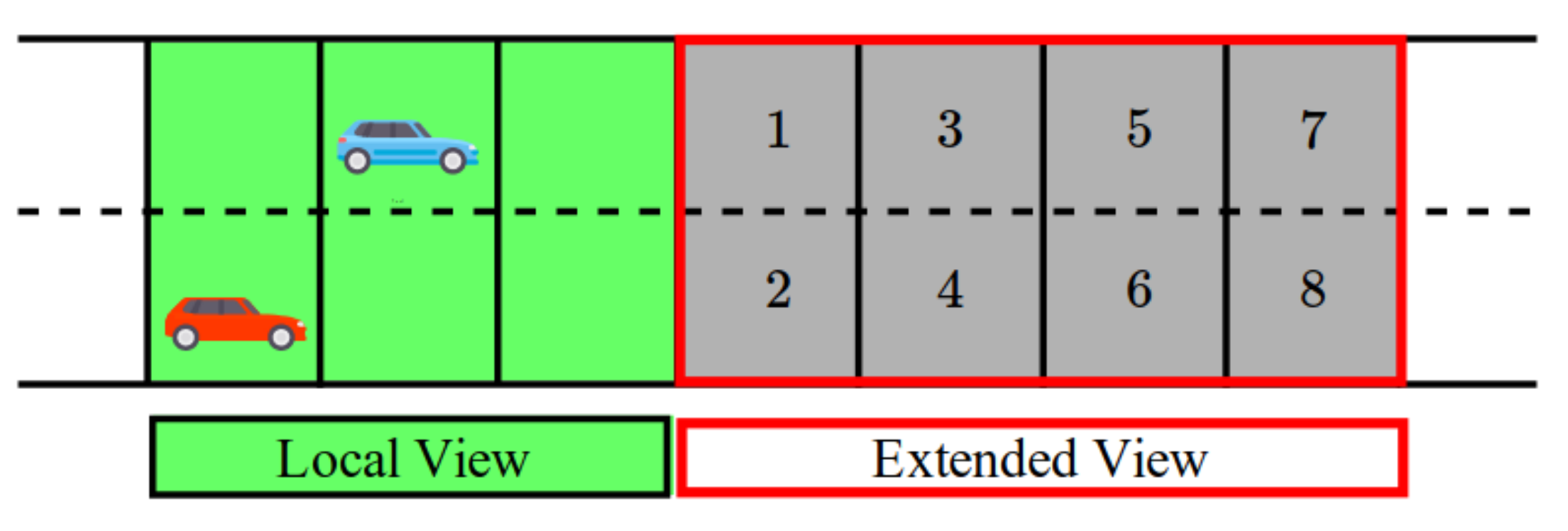}
		\caption{The grid maintained by the simulator. We show an instance in which the local view of the ego vehicle has a car.}
		\label{fig:evalView}
\end{figure}

\begin{figure*}[tb]
\begin{center}
\subfloat[Distance covered]{\includegraphics[height=0.25\textwidth]{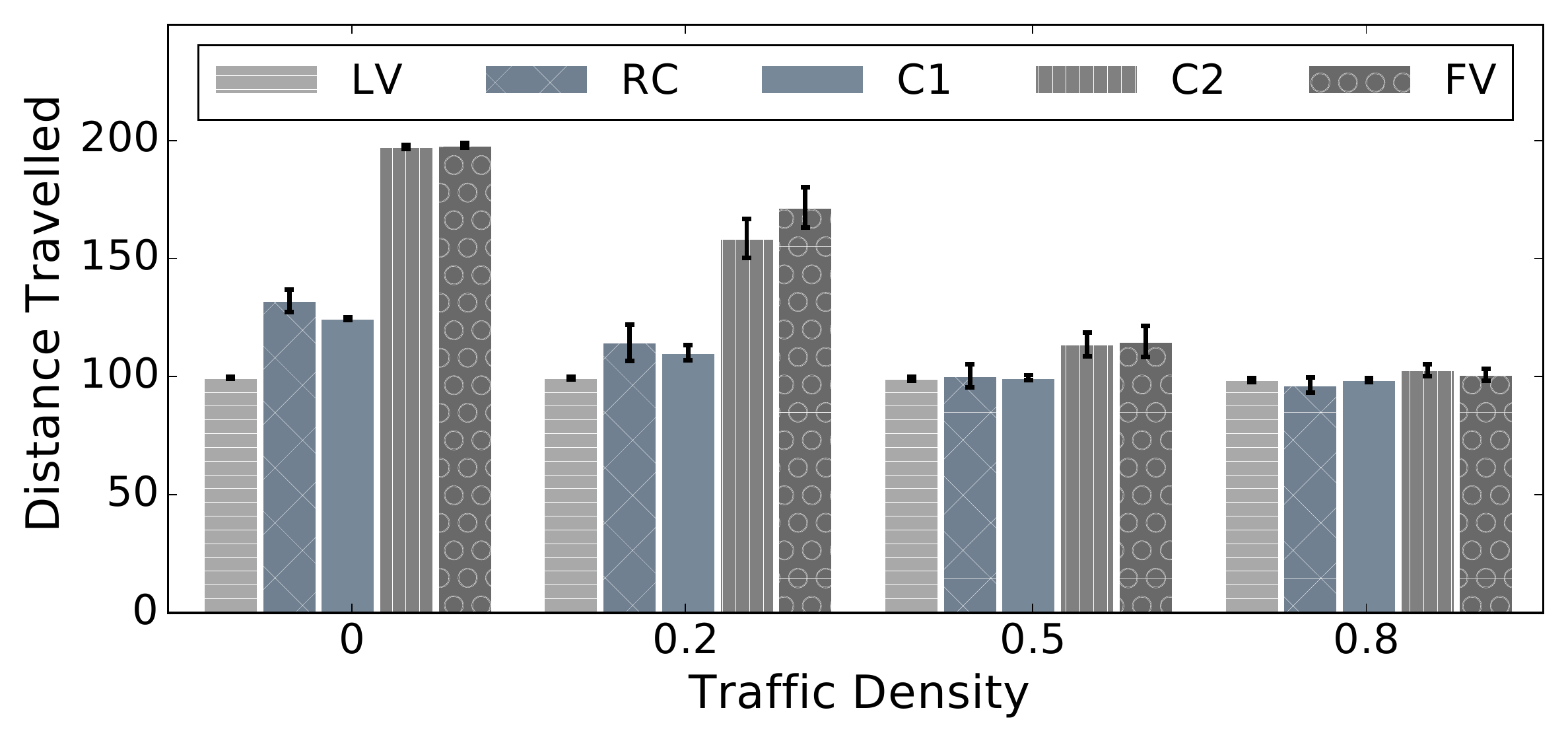}\label{fig:distbar}}
\subfloat[Velocity distribution]{\includegraphics[height=0.25\textwidth]{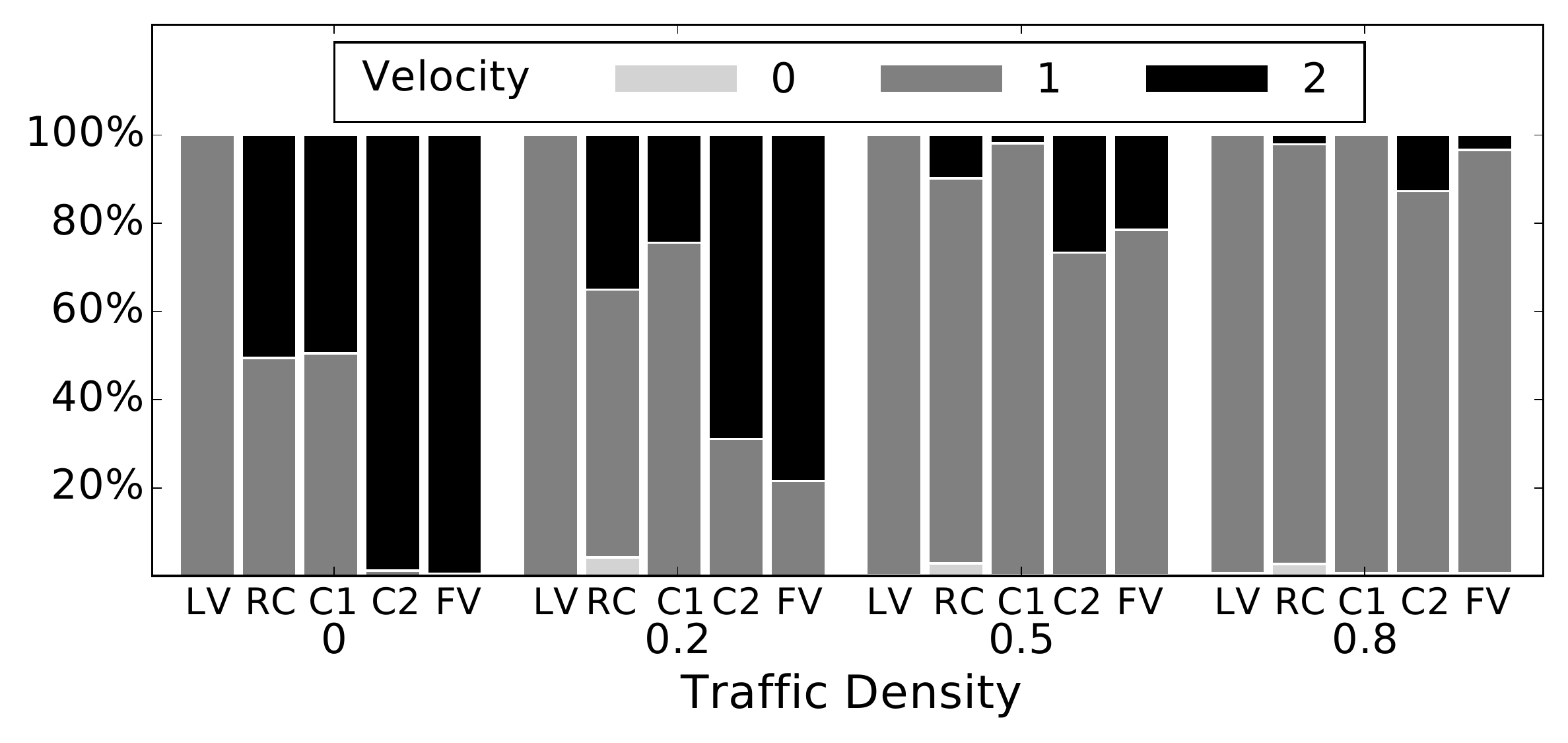}\label{fig:velbar}}\\
\subfloat[Motion planning action distribution]{\includegraphics[height=0.25\textwidth]{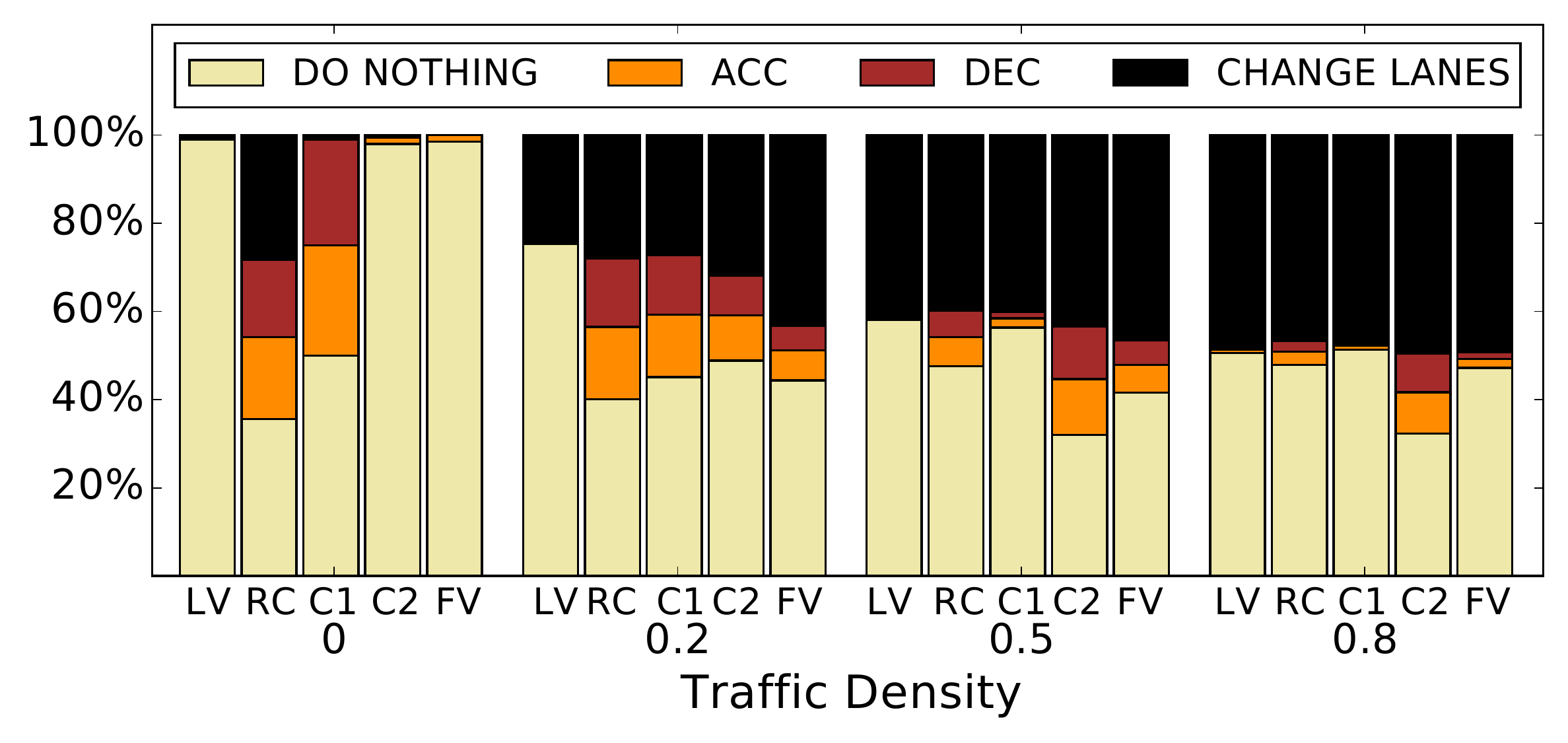}\label{fig:actionbar}}
\subfloat[Communications action distribution]{\includegraphics[height=0.25\textwidth]{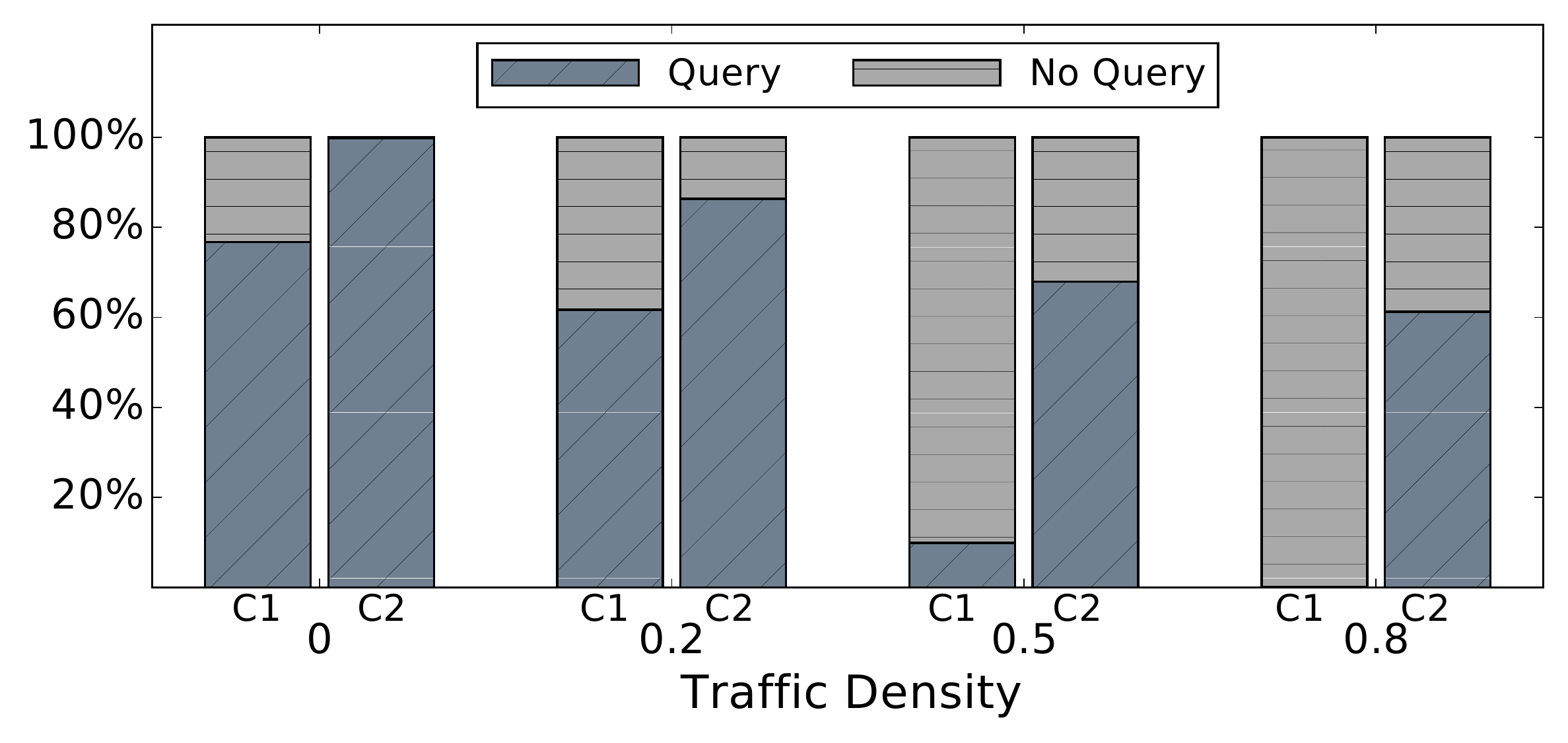}\label{fig:useOfComm2}}
\caption{Figures (a), (b), (c), and (d) show, respectively, the average distance covered, the distribution of the velocity of the agent, the distribution of the motion planning actions taken by the agent, and the distribution of the use of communications, calculated over $5000$ episodes of length $100$ time steps. These are shown for four different probabilities of occupancy (akin to traffic density). For each probability, we consider scenarios including Local view (LV), Random Communications (RC), Communications - one column at a time (C1), Communications - two columns at a time (C2), and Full View (FV).}
\end{center}
\label{steady_state}
\end{figure*}

We will demonstrate the efficacy of jointly adapting motion planning and communications actions in response to on-road and communications constraints. All results in this section were obtained for the grid shown in Figure~\ref{fig:evalView}. The ego vehicle could be in either lane, in its column of the grid. It was constrained to have a maximum velocity of $2$. It could choose an acceleration and deceleration of $1$ and $-1$, respectively. We will compare the following scenarios. For each of them, the optimal policy $\mu^*$, given by Equation~(\ref{eqn:mustarQ})), was obtained using Q-learning, as described earlier.
\begin{enumerate}[leftmargin=*]
\item Local View (LV) Only: The ego vehicle has no access to communications. It can only adapt its motion plan based on its local view.
\item Random Communication (RC): The ego vehicle receives information about its extended view from the infrastructure. At any time instant, it receives the occupancy of cells $1,2,3,4$ or that of cells $5,6,7,8$, independently of the previous time instants, and with equal probability. The vehicle chooses its motion planning actions given its local view and the occupancy received for the extended view.
\item Query of one column (two cells) at a time (C1): The ego vehicle may query the infrastructure to populate its extended view. A query at any time instant can be about exactly one of the four columns, the groups of cells $(1,2)$, $(3,4)$, $(5,6)$, and $(7,8)$. This implies a communications constraint of $2$ cells per time step. C1 has about $15\times 10^5$ state-action pairs for which Q-factors must be calculated.
\item  Query of two columns (four cells) at a time (C2): Similar to C1. However, exactly one of the two groups $(1,5,2,6)$ and $(3,7,4,8)$ may be queried by the ego vehicle. This amounts to having a communications rate of $4$ cells per time step, which is twice that available in C1. C2 has about $7\times 10^5$ state-action pairs.
\item Full View (FV): The ego vehicle obtains, without the need to query the infrastructure, the occupancy of both its local and extended views at any given time instant. This amounts to the ego vehicle having a large local view that includes all cells in the grid in Figure~\ref{fig:evalView}. FV has about $1.5 \times 10^5$ state-action pairs.
\end{enumerate}
Note that only in C1 and C2 does the ego vehicle jointly choose the motion planning and communications actions. In LV, RC, and FV, the ego vehicle chooses only a motion planning action. However, it has access to occupancy information that ranges from local only in LV to the entire grid in FV.

For each scenario, we \emph{test} the obtained policy over a set of $5000$ randomly chosen episodes, for a fixed choice of probability $pOccupied$ of a cell being occupied, where each episode was $100$ time steps long. We show results for $pOccupied = \{0, 0.2, 0.5, 0.8\}$. These choices correspond to on-road environments varying from no occupancy (no traffic) to very high occupancy (heavy traffic). During the episodes we ensure that both the cells in adjacent lanes are never occupied. This ensures that the ego vehicle has a free cell to move into for all the $100$ time steps in an episode. Recall from Algorithm~\ref{alg:qlearning} that, for any scenario, the policy is obtained by randomly choosing the value of $pOccupied$. Specifically, we don't train the ego vehicle separately for the different values of $pOccupied = \{0, 0.2, 0.5, 0.8\}$.

\emph{Understanding gains from an extended view:} Figure~\ref{fig:distbar} shows the distance traveled by the ego vehicle for different traffic densities and the scenarios we detailed above. Not surprisingly, the distance covered in FV is at least as much and often better than that covered under the other scenarios. On the other hand, given the limited view that the ego vehicle has in LV, the distance covered is smaller than the rest. The ego vehicle does almost as well in C2 as in FV. This is because the policy in C2 is able to query occupancy of all the cells in the extended view (Figure~\ref{fig:evalView}) in two time steps (four cells (two columns) at a time). Given that the maximum speed of the ego vehicle is $2$, C2 is in effect the same as FV.

Observe that the differences in distances covered under the different scenarios shrink as the traffic density increases. Also, in Figure~\ref{fig:velbar}, observe that as traffic density increases the agent chooses a velocity of $1$ more often, in all the scenarios. These two effects are in fact related. At large traffic densities the ego vehicle must restrict its speed to $1$ to avoid collisions even if it knows the occupancy of all cells in the grid. Note that at a velocity of $2$ the ego vehicle moves two cells, either in the same lane or in an adjacent lane, in one time step. While at any given time step, the ego vehicle will find an empty cell to enter, the probability of finding two consecutive empty cells is small for large traffic densities. To summarize, larger traffic densities constrain the velocities that the motion plan can choose. As a result, the knowledge of occupancy of cells in the extended view brings smaller rewards for larger densities.

\emph{Adapting communications to a constrained on-road environment:} While in FV the ego vehicle gets the entire view without any communications, one would hope that C1 and C2 would query less in high density scenarios as they jointly adapt motion planning and communications. This is in fact the case and can be seen in Figure~\ref{fig:useOfComm2}. Note that as the traffic density increases, the fraction of times No Query is used as a communications action increases. It is about $40\%$ for C2 at a density of $0.8$. On the other hand, C1 does not query at all at the density of $0.8$. This is because, given that it can query only two cells (one column) at a time, it is never able to benefit from the resulting extended view and it always has the ego vehicle move at a velocity of $1$ (Figure~\ref{fig:velbar}).

\emph{Adapting motion planning actions to communications constraints:}
From Figure~\ref{fig:distbar}, we can see that in C2 the ego vehicle covers the largest distance amongst RC, C1, and C2. In RC the ego vehicle has no control over the received information. However, it receives $4$ cells per time step as described earlier. This is the same as what the ego vehicle can query in C2. Clearly, from Figure~\ref{fig:distbar}, there are huge gains to choosing the communications action smartly. On the other hand, communications in C1 are constrained because of a smaller available communications rate ($2$ cells (one column) per time step). The impact of this on motion planning is apparent on seeing the distributions of velocity (Figure~\ref{fig:velbar}) chosen by the ego vehicle in C1 and C2. The constrained communications in C1 restricts the fraction of time the ego vehicle chooses the maximum velocity of $2$. The differences are especially significant for low traffic densities, that is when the on-road environment does not constrain motion planning.

Observe from Figure~\ref{fig:distbar} that RC shows gains over C1 for lower densities and does as well for higher densities. The gains are because in RC the communications rate is twice that in C1. This larger rate compensates for the randomness regarding what information must be sent in RC.

\emph{Action Selection:} Consider Figure~\ref{fig:actionbar}, which shows how motion planning actions are distributed across the scenarios and for different traffic densities. For a traffic density of $0$, LV, C2, and FV almost always choose Do Nothing. This is because, given the absence of any other occupants, in LV the ego vehicle sticks to a velocity of $1$ and in C2 and FV it sticks to a velocity of $2$. C1 and RC try to opportunistically benefit from communications. When the ego vehicle has a large enough extended view, it accelerates. At a higher velocity, it covers distance faster and isn't able to acquire the required extended view to stay fast. Hence, it decelerates. For similar reasons, at higher traffic densities, C2 sees a lot more of acceleration and deceleration than FV. In C2, the ego vehicle opportunistically increases velocity when communications reveals space in the extended view. However, this also means that it is forced to decelerate when the density is high. In FV, on the other hand, the ego vehicle is more often at a velocity of $1$ (see Figure~\ref{fig:velbar} for densities $0.5$ and $0.8$). Both C2 and FV see similar distance rewards (Figure~\ref{fig:distbar}), however.

Finally, observe that as traffic density increases the frequency of lane changes increases. This is simply because the ego vehicle encounters occupied cells more often.

%% file: conclusions.tex
\section{Conclusions and Future Work}
We formulated a reinforcement learning problem in which an autonomous vehicle jointly chooses a motion planning action and a communications action at every time step, such that its driving utility is optimized. We used the Q-learning algorithm to make the ego vehicle learn the optimal policy using simulations. We demonstrated how jointly adapting motion planning and communications actions allows an autonomous vehicle to (a) make judicious use of the communications network in an on-road environment that constrains feasible motion plans and (b) smartly choose motion planning given communications constraints. 

In the future we plan to investigate more realistic motion and uncertainty models and leverage \emph{deep} reinforcement learning techniques to obtain policies. In addition, extensions to networks of autonomous vehicles are planned.
\label{sec:conclusion}